\documentclass[%
 reprint,
 amsmath,amssymb,
 aps,
]{revtex4-2}
\usepackage[latin1]{inputenc}
\usepackage[active]{srcltx}
\usepackage{setspace}
\usepackage{graphicx}
\usepackage{dcolumn}
\usepackage{bm}
\usepackage{epsfig}
\usepackage{epstopdf}
\usepackage[arrowdel]{physics}
\usepackage{caption}
\usepackage{subcaption}
\usepackage{chngcntr}

\usepackage{pgfplots}

\begin{document}

\preprint{DFIST/2021.1}

\title{On Electromagnetic Turbulence}

\author{Mario J. Pinheiro}
\affiliation{ Department of Physics, Instituto Superior T\'{e}cnico-IST, Universidade de Lisboa-UL, Av. Rovisco Pais, 1049-001 Cedex, Lisboa, Portugal 
}%




\date{\today}

\begin{abstract}
Combining a generalized current in the set of Maxwell's equations offers a useful framework to address the complex phenomena of electromagnetic turbulence. The fluidic-electromagnetic analogy implies that diffraction is the analog phenomenon of EM turbulence and indicates norms to design suitable plasmon circuity to control electromagnetic turbulence in stealth technology and propulsion machines.
\end{abstract}

\maketitle


\section{\label{sec:level1}Introduction}

In 1883 Osborne Reynolds discovered the phenomenon of fluid dynamic
turbulence when studying the flow of water in a cylindrical pipe
driven by a pressure gradient. He found that when a critical
velocity (well characterized by a critical value of the so-called
Reynolds number $Re_c$) was exceeded, the flow becomes turbulent.
However, it persists an absence of an adequate understanding of the origin of turbulence, outlasting a fundamental challenge to scientists and engineers as well, considering that most significant flows are turbulent.

Large-scale computational, and experimental capabilities at disposal of researchers and engineers certainly help in better comprehension and managing the source of turbulent flows.
It is comprehended that turbulence is a random solenoidal motion of the
fluid accompanied by a large increase of transport properties, such as
viscosity (momentum), diffusivity (mass), heat conductivity
(energy), and resistivity (electric current). The turbulent flow is
energetically fed by the main flow, and energy losses may rise
as a pressure drop or friction loss (with energy spectra $E(k)=C_K \epsilon^{2/3}k^{-5/3}$, according to Kolmogorov's theory). Turbulence also occurs in
conducting fluids (ionic flows) due to an interaction with the background 
electromagnetic field (radial profile of $\pmb{E} \times \pmb{B}$ shear flows at the edge of fusion device characterized by the gyrocenter shift induced by the collisions between ion and neutral~\cite{Navarro_2020}). For an inviscid, ideally conducting fluid in
the presence of an electromagnetic field, the magnetic field lines
hook to the fluid. If the fluid has a small resistivity, then
the magnetic field lines will slowly diffuse through the medium. If turbulence may be defined as the field of random or chaotic vorticity, then the noise may be defined as the random motion of boundaries~\cite{Saffman}. Although there are some difficulties in defining turbulence, it seems to exist a consensus that turbulence evolution in a fluid is a thermodynamically irreversible process~\cite{Kiehn_1}. Furthermore, according to von Karman~\cite{Rao_2010} turbulence appears when fluid flows past solid surfaces or by the flow of layers of fluids with different velocities past or over one another.
Marmanis suggest that vorticity ($\mathbf{w}=\boldsymbol{\nabla} \times \mathbf{u}$) and the Lamb vector ($\mathbf{l}=\mathbf{w}\times \mathbf{u}$) should be central to the theory of turbulence~\cite{Marmanis_1998}.

In this article, we aim to contribute for physical insight on the electromagnetic turbulence from the standpoint of classical fields, supported on established analogies with turbulent hydrodynamics, namely by giving evidence that optical diffraction is the analog of fluid turbulence, means to control EM turbulence and shielding, and explore in some detail the nature and applications of this coupling. 

\subsection{\label{sec:level2}Modification of Maxwell's equations}

Maxwell put the generalized current under the general form (as it follows from the Helmholtz's theorem):
\begin{equation}\label{eq1}
\mathbf{C}=\mathbf{J}+\frac{\partial \mathbf{D}}{\partial t} + \mathrm{curl} \; \mathbf{Z},
\end{equation}
with $\boldsymbol{\nabla} \cdot \mathbf{C}= 0$, without specifying, or going further, on the characterization of the physical quantity $Z$.  We can verify that this new field has the dimension of a surface density of current $[Z] =A/m$. Eq. 1 encodes the Hodge-de Rham decomposition theorem for a regular p-form of degree p is a sum of exact, harmonic, and co-exact p forms.

Vorticity generates turbulence, and Stoke's theorem indicates that a local vortex pinned at a location will drag fluid in a rotating state, even in the case of a perfect fluid~\cite{Lesieur_94}. This effect
is analog to the production of a magnetic field by a rectilinear
current, a sort of {\it induction of velocity}. Vorticity is the seed of turbulence.
The vortex starts to grow on a mixed layer of magnitude
$\delta (y)$ where $x$ is the distance relative to the point where
the flowing starts, the leading edge. Consequently, we can assimilate the vortex as a kind-of surface molecular current or eddy current. Eddy currents are dissipative processes.
We will represent the structure of this vector field by the {\it Ansatz} $\mathbf{M}$, such as an associated induced current density $\mathbf{J}_{rot}=\mathrm{curl} \:\:\mathbf{M}$, is created, introducing the magnetization vector field $\mathbf{M}$ (with $\mathbf{J}_{rot}$ in units A$/$m$^2$).

But it exists a well-known relationship between the magnetization vector and the magnetic field intensity $\mathbf{H}$, $\mathbf{M}=\chi_m \mathbf{H}$, where $\chi_m$ is a dimensionless quantity called magnetic susceptibility, and $\mathbf{H}=\frac{1}{\mu}\mathbf{B}$, then we may write
\begin{equation}\label{eq2}
\mathbf{J}_{rot}=-\frac{\chi_m}{\mu}\Delta \mathbf{A},
\end{equation}
since $\pmb{\nabla} \cdot \mathbf{A}=0$, ensuring continuity in the current flow and $\mathbf{B}=\pmb{\nabla} \times \mathbf{A}$. It
results from the above that the Amp\`{e}re's equation should read
instead:
\begin{equation}\label{eq3}
c^2[\pmb{\nabla} \times \mathbf{B}] = \frac{\partial \mathbf{D}}{\partial t} + \mathbf{J} - \frac{\chi_m}{\mu} \Delta
\mathbf{A}.
\end{equation}
The previous Eq.~\ref{eq3} appears (without the new term) more frequently under the form of Amp\`{e}re's equation for vector $\mathbf{H}$:
\begin{equation}\label{eq3aa}
\frac{\partial \mathbf{D}}{\partial t}= [\nabla \times \mathbf{H}] -
\mathbf{J},
\end{equation}
after introducing a new excitation magnetic field $\mathbf{H}\equiv
\mathbf{B}/\mu_0 - \mathbf{M}$. Eq.~\ref{eq3} has a dissipative
component in the last term. It allows a straightforward deduction
of London's equation characterizing superconductors. Inside the
sample $\mathbf{B}=0$ and as $\mathbf{E}$ arises from a changing
magnetic field, then $\mathrm{grad} \: V=0$. So, Eq.~\ref{eq3} gives straightforwardly
\begin{equation}\label{eq3a}
    \nabla^2 \mathbf{A}-\frac{\mu}{\chi_m}\mathbf{J}=0.
\end{equation}


London admitted that the superconductor consisted in a condensation
in a state of zero momentum, and hence:
\begin{equation}\label{eq3b}
    \mathbf{P}=m\mathbf{v}+q\mathbf{A}=0.
\end{equation}
But since $\mathbf{J}=nq\mathbf{v}$, then, from Eq.~\ref{eq3b} we obtain
\begin{equation}\label{eq3c}
    \frac{m}{nq^2}\mathbf{J}+\mathbf{A}=0.
\end{equation}
Combining Eq.~\ref{eq3a}-Eq.~\ref{eq3c} and noting that for a superconductor, $\chi_m=-1$, we obtain:
\begin{equation}\label{eq3d}
    \nabla^2 \mathbf{A}-\lambda_L^2\mathbf{A}=0,
\end{equation}
admitting the solution $\mathbf{A}=\mathbf{A_0} \exp(-z/\lambda_L)$,
with $\lambda_l^2=\mu nq^2/m\chi_m$ denoting the London penetration
depth, the characteristic distance over which the field penetrates the superconductor. $q$ is the effective charge of the carriers of the superconducting state (i.e., $q=2e$, with $-e$ denoting the electron charge).

As shown in previous work~\cite{Pinheiro_09} the
electromotive force is given by:
\begin{equation}\label{eq4}
\rho\mathbf{E}=-\rho \nabla \phi - \rho \frac{D \mathbf{A}}{Dt}.
\end{equation}
Adding the term corresponding to the vortex structure, as subsumed in Eq.~\ref{eq1}, we introduce 
\begin{equation}\label{eq5}
\rho \mathbf{E}=-\rho \nabla \phi - \rho \frac{\partial
\mathbf{A}}{\partial t} - \rho \nabla (\mathbf{v} \cdot \mathbf{A})
+ \rho [\mathbf{v} \times \mathbf{B}] - n_s\frac{\hbar}{\mid e \mid} \pmb{\nabla} \times \mathbf{M}.
\end{equation}
The last term on the r.h.s. contains two fundamental constants, $\hbar$ and $e$, to secure homogeneity, and for consistency with the quantization of the magnetic flux. Notice that in superconductors, we have the flux given by $\Phi$ quantized according to $\frac{e \Phi}{\hbar c}=2 \pi n$, $(n=0,\pm1, \pm2,...)$, with $n_s$ being the surface density of vortices filling the texture on the boundary layer above the blunt-body. The rationale of this simple choice is confirmed when compared with known phenomena.
Using the relations $\mathbf{M}=\chi_m \mathbf{H}$ and
$\mathbf{H}=\frac{1}{\mu} \mathrm{curl}\:\: \mathbf{A}$, we finally obtain:
\begin{equation}\label{eq6}
\frac{\partial \mathbf{A}}{\partial t}= \frac{\mathbf{F}^{ext}}{\rho} - \nabla \phi - \nabla (\mathbf{v} \cdot \mathbf{A}) + [\mathbf{v} \times \mathbf{B}] + \nu_{em} \Delta \mathbf{A}.
\end{equation}

\subsection{The electromagnetic viscosity term}

From the above Eq.~\ref{eq6} we are required to define
\begin{equation}\label{eq6a}
\nu_{em} \equiv \frac{\hbar}{\rho} \left( \frac{n_s \chi_m}{|e|\mu} \right),
\end{equation}
that plays the role of an {\it electromagnetic viscosity}, provided we replace $\mathbf{E}=[\mathbf{v} \times \mathbf{B}]$, a result also proposed by Marmanis in his Thesis~\cite{Marmanis}.

The introduction of the electromotive force in connection with the
convective (total) derivative drives us directly to the specific
condition, which must be satisfied by $\mathbf{B}$. Taking the
rotational of Eq.~\ref{eq6} and not regarding external forces, is straightforwardly taken the following expression
(e.g.,~\cite{Batchelor}):
\begin{equation}\label{eq7}
\frac{\partial \mathbf{B}}{\partial t} = \mathrm{curl} \:\:[\mathbf{v} \times
\mathbf{B}] + \eta \Delta \mathbf{B},
\end{equation}
where $\eta$ is the resistivity of the medium, and it is equivalent to the magnitude obtained earlier, that is, $\eta \equiv \nu_{em}$. Eq.~\ref{eq7} is the induction equation for the magnetic field in magnetohydrodynamics. In MHD this is taken to be $\eta=m_e <\nu_{ei}>/n_e
e^2$ in SI units, where $\mu_0$ is the magnetic permeability. In order for the analogy to be completed we have $\eta \equiv n_s \hbar/e \rho$, with $\rho=ne$, or

\begin{equation}\label{eq7a}
\eta = \frac{n_s}{n_v}\frac{\hbar}{e^2}\chi_m.
\end{equation}
This result attests to the nature of the electric resistivity depending on the prefactor $n_s/n_v$, which shows its dependency from the electrons surface density, that without electrons present on a sheath-layer, no electromagnetic turbulence is achievable. Still, Eq.~\ref{eq7a} must be valid when analyzing an electromagnetic wave incident to an obstacle, when light passes through an aperture and the formed diffraction pattern.

Using the Bohm sheath criterion were ions are supposed to leave the sheath with speed $u_s$ above the Bohm speed $u_B=\sqrt{\frac{ek_BT_e}{M}}$, such as $u_s \geq u_B$, and considering that the potential drop $\Phi_p$ across the sheath-presheath launch the ions until they reach the Bohm speed, $Mu_B^2/2=e\Phi_p$, we have $\Phi_p=T_e/2$, delivering to us a thumb rule for the ratio of the density at the sheath edge $n_s$ over the density in the plasma $n_v$, such as $n_s \approx \frac{25}{41}n_v$ (see, e.g., Ref.~\cite{Lieberman}). Hence, in normal conditions, we may expect that the resistivity of the medium may have a maximum value:
\begin{equation}\label{eq7b}
\eta \leq \frac{n_s}{n_v}\frac{\hbar}{e^2} \chi_m.
\end{equation}
The Eq.~\ref{eq7b} implies that higher magnetic susceptibilities $\chi_m$ contribute to a substantial increase of the electric resistivity of the medium, typical in ferrimagnetic substances exhibiting high values of $\chi_m$, and high electrical resistivity. Additionally, insulants, only possessing more increased surface charge density than volume charge, can hold high electric resistivity.

Conversely, Eq.~\ref{eq7a} gives for the surface density:
\begin{equation}\label{}
n_s =\frac{m_e <\nu_{ei}> \mu}{\hbar \chi_m}.
\end{equation}
In a typical plasma we have $<\nu_{ei}>\approx 10^9$ s$^{-1}$ and
for electrons we got $n_s \approx 10^3/\chi_m$ particles$/$cm$^2$. For example, for hydrogen $\chi_m \approx 10^{-9}$ and $n_s \approx 10^{12}$ electrons per cm$^2$. The obtained expression for the electromagnetic viscosity scales well with the (magnetic diffusion based) Sweet-Parker sheet-model, $\tau =L^2/\eta$, with $L$ representing a field diffusion region, for example, $L \approx 10^4$ km in a solar flare~\cite{Biskamp_2005}. Our estimated value for $\chi_m$ gives a too much longer time of reconnection in a solar flare, typically of the order of tens of minutes. However, a possible discrepancy may have originated on the physical mechanism related that emerges from the presence of the Planck's constant $\hbar$ and the magnetic susceptibility $\chi_m$ in Eq.~\ref{eq7b}.

Therefore, we may consider Eq.~\ref{eq6} and, using the fluidic electrodynamics analogy, write the Navier-Stokes equation:
\begin{equation}\label{}
\frac{\partial \mathbf{u}}{\partial t}=-\nabla p -(\mathbf{u} \cdot \nabla \mathbf{u}) + \nu \Delta \mathbf{u}.
\end{equation}

\subsection{Time rate of magnetic reconnection}

The preceding arguments lead to the construction of a new expression for the viscosity, in terms of superficial density of fluid, and bulk density, as the critical ratio.
One immediate advantage attained with this model stays in the important separation between the two groups of electrons, let us say, the normal (bulk) electrons $n_v$ and the surface electrons $n_s$ (that emerge at the surface of the medium, in the plasma sheet), and the total density of conduction electrons is $n=n_s+n_v$.
Eq.~\ref{eq7} provide us with a new formula for the time decay for a magnetic field (diffusion time):
\begin{equation}\label{eq7c}
\tau \simeq \frac{n_v}{n_s}\frac{e^2 L^2 \mu}{\hbar \chi_m}.
\end{equation}
Here, $L$ is a global length scale of change in the magnetic field. From Eq.~\ref{eq7c} we obtain an estimate for the time rate of magnetic reconnection, $\tau \approx 10^9$ years, assuming $n_v=10^{12}$ electrons per cubic centimeter, and $n_s=0.1 n_v$. However, the term $\chi_m$ may become negative (and growing in magnitude) in an ideal plasma dominated by the diamagnetic effect of the charged particles in their helical trajectory around the magnetic field lines (see, e.g., Ref.~\cite{Lowes_2007}). As the forward Eq.~\ref{eq8} shows, for a negative value of $\chi_m$ a spontaneous instability appears leading to $|\pmb{A}| \sim e^{\eta k^2 t}$ for wavenumber $k$. Now, a plasma in pressure-balance equilibrium is perfectly diamagnetic (see, e.g., Ref.~\cite{Schmidt_1966,Bhattacharjee_2020}), a state of generalized zero vorticity and helicity, 
the ends of the loop anchored in the dense photosphere, and this is the first state of minimum energy before the kink instability surging from a highly twisted flux rope evolve to a coronal mass ejection~\cite{Wang_2017}, the large eruption of plasma and magnetic field from a star leading to the destruction of excess energy.

\subsection{\label{sec:citeref}Spiral structures and turbulence}

Eq.~\ref{eq7} can be compared with the dynamical equations describing the fields the vorticity $\omega$ and Lamb vector $\mathbf{l}$:
\begin{equation}\label{eq7aa}
\frac{\partial \mathbf{\omega}}{\partial t} = -[\nabla \times \mathbf{l}] + \nu \nabla^2 \mathbf{\omega}
\end{equation}
and as well
\begin{equation}\label{eq7bb}
\nabla \cdot \mathbf{\omega} = 0.
\end{equation}
can be the analogue of $\nabla \cdot \mathbf{B}=0$. Eq.~\ref{eq7aa} is equivalent to the equation of conservation of angular momentum~\cite{Chatwin_1973}. It can be shown that the parallel component of the angular momentum is conserved in a fully developed turbulence, result encoded in the Loitsyansky-like integral $I_{\parallel}=-\int r_{\perp}^2 < \mathbf{u}_{\perp} \cdot \mathbf{u}^{'}_{\perp} > d \mathbf{r}$, with $<\mathbf{u} \cdot \mathbf{u}^{'}>$ denoting the usual two-point velocity correlation and $\perp$ indicating the component perpendicular to the external magnetic field~\cite{Okamoto_2020}. It is a cornerstone in treating turbulence.

Eq.~\ref{eq6} results in a diffusion type equation for each component of the vector potential field $\mathbf{A}$ (in a rest-frame, $\mathbf{v}=0$) that come out from the modified Amp\`{e}re's equation~\ref{eq3aa}:
\begin{equation}\label{eq8}
\frac{\partial \mathbf{A}}{\partial t}=\nu_{em} \nabla^2 \mathbf{A} + \mathbf{s(\mathbf{r},t)}.
\end{equation}
The theory of turbulence in fluids is primarily founded on the flow representation of the vorticity field. Hence, we expect from Eq.21 to obtain a valid representation of the EM turbulence.
We know that diffraction through an aperture can be explained
through Faraday's induction law~\cite{Zolotorev_00}, using Eq.~\ref{eq7}.
We propose here an additional vectorial source term $\mathbf{s(\mathbf{r},t)}$, which might represent an incident electromagnetic wave hitting an obstacle. Remark that there is a field penetration depth, $\lambda_L=\sqrt{\frac{\nu_{em}}{f}}$, correlated to an electromagnetic wave with frequency $f$ which typically fall in the nanoscale (near-field) region. We intend to establish an analogy between both hydrodynamic and EM fields to solve problems in both fields, recommended by the similarity between Eqs.~\ref{eq7aa} and Eq.~\ref{eq8}.


From a parabolic partial differential equation, like Eq.~\ref{eq8}, we can obtain a hyperbolic partial differential equation applying directly the following field dependency
\begin{equation}\label{eq9}
\mathbf{A}(x,y,z,t) = \mathbf{A}(x,y,z)T(t),
\end{equation}
which is convenient for individual radiation sources, and obtaining a Sturm-Liouville type of equation where electromagnetic theory match geometrical optics, according to the Sommerfeld and Runge approach~\cite{Kline}
\begin{equation}\label{eq10}
\nabla^2 \mathbf{A}(\mathbf{r})+ k^2 \mathbf{A}(\mathbf{r})=\delta(\mathbf{r}).
\end{equation}
Here, $k=\sqrt{\epsilon \mu} \omega= 2\pi/ \lambda$ is a constant in a given medium, and, for consistency with the above described framework, we should put $\omega=\nu_{em} k^2$ (as noticed before, $\nu_{em}$ has the dimension $[L]^2/[T]$).

\subsection{Controlled electromagnetic turbulence}

Hence, the framework outlined above suggests that the solution of Eq.~\ref{eq10} represents the turbulent vector potential and that diffraction is the analog phenomena of EM turbulence. Besides, it shows that the surface density of plasmons present on the obstacle to the waves impacts the diffraction phenomena. The solution of Eq.~\ref{eq10} is given by (see Ref.~\cite{Marathay})
\begin{equation}\label{eq11}
\begin{split}
\mathbf{A}(\mathbf{r})=\frac{1}{4 \pi} \iiint_{\infty} [\mu_0 \mathbf{J}_{\bot e}(\mathbf{r}')+ \\
\epsilon_0 \mathbf{J}_{\bot m}(\mathbf{r}')]\frac{e^{i \sqrt{\frac{\omega}{\nu_{em}}} |\mathbf{r}-\mathbf{r}'|}}{|\mathbf{r}-\mathbf{r}'|}d^3 \mathbf{r}',
\end{split}
\end{equation}
where $\mathbf{A}(\mathbf{r},t))=\mathbf{A}(\mathbf{r})e^{i \omega t}$, and $\mathbf{J}_{\bot m,e}(\mathbf{r}')$ represent the transversal (electric and magnetic) current densities.
It points out the action of scalar and vectorial waves on diffraction effects when an electromagnetic wave interacts with
a surface layer of a material~\cite{Fujii_2016} (surface plasmons) or plasma, and to the possibility to control diffraction, aiming, for example, to enhance its intensity through gratings, or stealth technology, utilizing an appropriate combination of material properties and the light wave incident on its surface. Radiating EM waves are feed only by the transverse part of
the current density, the source of the vector potential.

Could an array change its reflectivity by optimally grading the density of surface plasmons? This feature is feasible if conditions i) $k \gg 1/R$, $R \gg \lambda$, and, ii) $z \approx R$, holds, then we rewrite for the plan $z>0$
\begin{equation}\label{eq11a}
\begin{split}
\mathbf{A}(x,y,z)=\frac{1}{4\pi} \iiint_{\infty} [\mu_0 \mathbf{J}_{\bot e}(\mathbf{r}')+ 
\epsilon_0 \mathbf{J}_{\bot m}(\mathbf{r}')] \\ e^{i[u_1(x-x')+u_2(y-y')]}dx'dy'e^{iz\sqrt{k^2-u_1^2-u_2^2}}/R
\end{split}
\end{equation}
with $R=\sqrt{(x-x')^2+(y-y')^2+z^2}$. Eq.~\ref{eq11} is valid for a monochromatic wave, but using the superposition principle we can construct a more general solution of the wave equation which includes plane waves with different amplitudes, phases and directions. Let us denote by $\alpha$ and $\phi$ the direction of propagation of the wave, with $k_x=u_1=k\sin \alpha$ and $k_y=u_2= k \cos \alpha \sin \phi$, and assume that the plan $(x,y)$ is coplanar to the diffraction structure. Eq.~\ref{eq11a} becomes (for $z>0$):
\begin{equation}\label{eq12}
\begin{split}
\mathbf{A}(x,y,z)=\frac{1}{4\pi^2} \iiint_{\infty} \mathbf{g}(u_1,u_2) e^{i[u_1(x-x^{'})+u_2(y-y')]}\\e^{iz\sqrt{k^2-u_1^2-u_2^2}}du_1 du_2.
\end{split}
\end{equation}
setting the wave plan distributions of amplitudes and phases in all directions, and with
\begin{equation}\label{eq13}
    \begin{split}
        \mathbf{g}(u_1,u_2)=\iint_{-\infty}^{\infty} \mathbf{A}(x',y',0) e^{-i[u_1(x-x')+u_2(y-y'))}dx'dy'
    \end{split}
\end{equation}
 representing the field angular spectra, build on the transversal current densities, sit on the surface. Consider the relevant case when the field doesn't depend on coordinate y, such as $\mathbf{A}(x,y,z)=\mathbf{A}(x,0,z)$, where the source is located only along the Ox axis.

\begin{figure}
\includegraphics[scale=1.1]{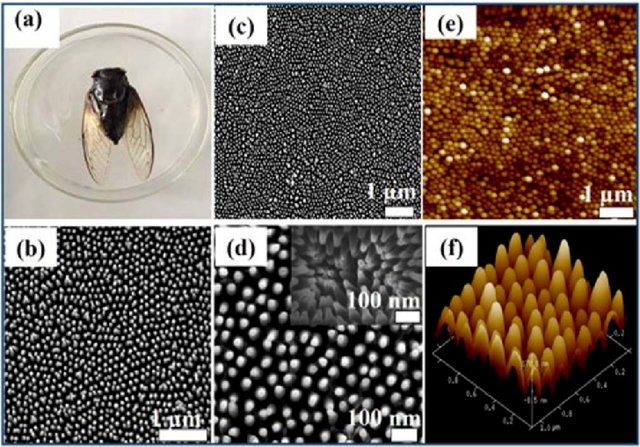}
\caption{Photograph of a black cicada wing. (b) and (c) Low magnified top-view SEM images of cicada wings. (d) Highly magnified top view SEM image of cicada wings [the side view SEM image of cicada wings is shown in the inset in (d)]. (e) The two-dimensional AFM image of the cicada wing. (f) The high-resolution three-dimensional AFM image of the cicada wing. Image courtesy: Ref.~\cite{Kim_2017}.}~\label{Fig2}
\end{figure}

Then
\begin{equation}\label{eq14}
\begin{split}
           g(u_1,u_2)=\iint \mathbf{A}(x',0) e^{i[u_1(x')+u_2(y')]}dx'dy'=\\2\pi \delta(u_2)g(u_1)
\end{split}
\end{equation}
with 
\begin{equation}\label{eq15}
    \mathbf{g}(u_1)=\int_{-\infty}^{\infty} \mathbf{A}(x',0)e^{iu_1(x')}dx'.
\end{equation}
The vector potential becomes:
\begin{equation}\label{eq16}
    \mathbf{A}(x,0,z)=\frac{1}{2\pi} \int_{-\infty}^{+\infty} g(u) e^{iz\sqrt{k-u^2}}e^{iux} du.
\end{equation}
Applying the method of stationary phase the vector potential can be written under the form (with $\mathbf{i}$ denoting the unit vector along the Ox axis, and as well $\mathbf{j}$, and $\mathbf{k}$):
\begin{equation}\label{eq17}
    \mathbf{A}(x,0,z)=\frac{1}{2 \pi}g(u_0) e^{if(u_0) \mp \frac{\pi}{4}}\sqrt{\frac{2\pi}{\mp f^{''}(u_0)}}e^{i \omega t} \mathbf{i}
\end{equation}
with
\begin{equation}
    f(u) = -z\sqrt{\frac{\omega}{\nu_{em}}-u^2}+ux.
\end{equation}
After multiplying by the independent temporal term (according to Eq.~\ref{eq9}, with 
\begin{equation}\label{eq18}
u_0=\pm \frac{x}{\sqrt{x^2+z^2}}\frac{\omega}{\nu_{em}}, 
\end{equation}
for the condition $f^{'}(u_0)=0$, and defining $f(u)$ such as:
\begin{equation}\label{eq19}
f(u_0) \equiv -z\sqrt{k^2-u_0^2}+u_0x.  
\end{equation}
The variation of the wave number $u$ that is related to phase variation of $f(u)$ from $0$ to $\pi$, is given by:
\begin{equation}\label{eq20}
    \Delta u= \sqrt{2 \pi /(\mp f''(u_0))}.
\end{equation}
The signs $\mp$ in front of $f^{''}(u_0)$ must be chosen for the argument of the square root to be positive, and $\mathbf{g}(u_0)$ is the angular spectrum of the wave field. 

Finally, Eq.~\ref{eq17} is written under the form:
\begin{equation}\label{eq21}
    \mathbf{A}(x,0,z)=\frac{1}{2 \pi} g(u_0) \sqrt{\frac{2\pi \omega}{\nu_{em}}} e^{i \omega t} \frac{z}{(x^2+z^2)^{3/4}} e^{\imath f(u_0) \mp  \frac{\pi}{4}} \mathbf{i}.
\end{equation}

Then the conditions for anti-reflectivity, or superluminosity, are obtained. The maximum and minimum intensity $A$ patterns are observed within the following conditions (where $n$ is an integer), $x_n$ expressing the position on the Ox axis of the $n$ radiation sources:
\begin{equation}
   \mathbf{A}(x,0,z) = \left\{
   \begin{array}{lr}
        A=0, & \text{for } x_n=\pm \pi (n+\frac{1}{2})\frac{\nu_{em}}{\omega},  \\
        \\
        A=A_{max}, & \text{for } x_n = \pm 2 \pi n \frac{\nu_{em}}{\omega}
        \end{array} 
        \right. 
\end{equation}

\begin{figure}
\includegraphics[scale=0.27]{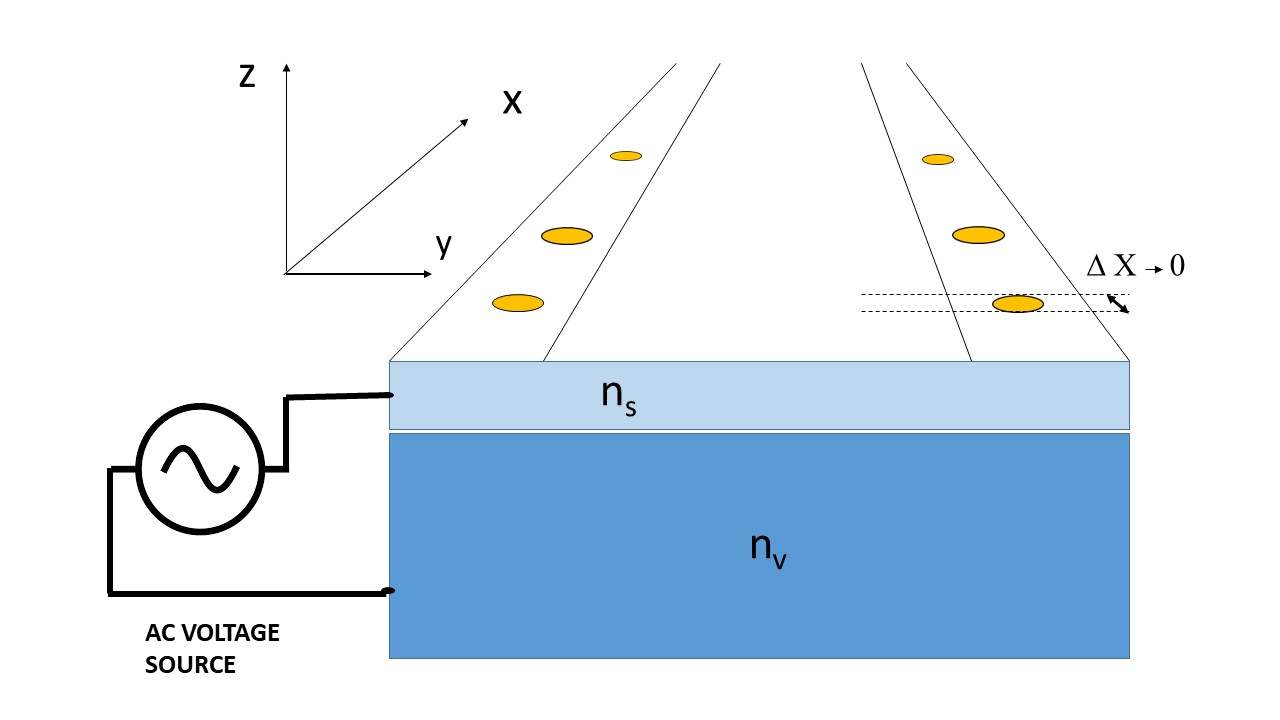}
\caption{Plasmon circuit to induce electromagnetic amplification or attenuation. In the Fraunhofer diffraction (in the far-field region) analyzed here, in the far-field region, $\Delta x$ determines the possible extent of the source, or pillar, as in cicada wings, here discussed as a punctual source.}~\label{Fig3}
\end{figure}

The electric field can be calculated, giving
\begin{equation}
    \mathbf{E}=\frac{\partial \mathbf{A}}{\partial t}=\frac{1}{2\pi}g(u_0) i \omega e^{\imath \omega t} \sqrt{\frac{2 \pi \omega}{\nu_{em}}}e^{\imath f(u_0) \mp \frac{\pi}{4}}\frac{z}{(x^2+z^2)^{3/4}} \mathbf{i},
\end{equation}
and the magnetic field is
\begin{equation}
    \begin{split}\mathbf{B}=\nabla \times \mathbf{A}=\frac{1}{2 \pi}g(u_0)e^{i\omega t} \sqrt{\frac{2 \pi \omega}{\nu_{em}}}e^{\imath f(u_0) \mp  \frac{\pi}{4}}\\ \left[ \frac{x^2(2-2i\sqrt{\frac{\omega}{\nu_{em}}}\frac{xz}{\sqrt{x^2+z^2}})+z^2(-1-2i\sqrt{\frac{\omega}{\nu_{em}}}\frac{xz}{\sqrt{x^2+z^2}})}{2(x^2+z^2)^{7/4}} \right]\mathbf{k}.
\end{split}
\end{equation}

Figs.~\ref{plasmons1}-~\ref{plasmons2} illustrate the effect of an applied electromagnetic wave of frequency $\omega$ acting on plasmons present on one point-like nanodevice, with one surface layer with $n_s$ surface electrons density, sit on a metallic surface, or plasma, with $n_v$ volumetric electron density. An EM wave with frequency $\omega$ actuate on electrons that shine modulated light above and below the nanodevice. It is assumed $z \ll 1$ in the numerical calculations shown in Figs.~\ref{plasmons1}-~\ref{plasmons4}, representing the intensity of the emitted light $I =<|\mathbf{E} \cross \mathbf{H}|>$, with the Poynting vector, $\mathbf{S}=\frac{1}{\mu}\mathbf{E} \cross \mathbf{B}$, including only the real part of the fields. We may notice that the Poynting vector aligns along the vertical axis, despite the EM momentum confined to a restricted volumetric region, with a maximum adjacent to the surface. The intensity distribution inside the thin surface layer where the EM viscosity is build-up, is due to the plasmons vortex generating the effect of surface EM waves, that develop in multiple beam-like structures, depending on the ratio $\omega/\nu_{em}$, as portrayed in Figs.~\ref{plasmons3}-~\ref{plasmons4}, with evidence of turbulent, classical fractal-like behavior~\cite{Sreenivasan_Meneveau}, when $\omega/\nu_{em} \gg 1$, with similarity to a transition to turbulence in boundary layers. The scattered waves do not progress isotropically but are transmitted or reflected, bordering along with the device structure. Similar streamwise vorticity fluctuations induced by near-wall vortices are visible in tubular structures in plane channel flows~\cite{Dubief_2000}.

    \renewcommand{\thefigure}{3}
\begin{figure}
     \begin{subfigure}[b]{0.35\textwidth}
         \includegraphics[width=\textwidth]{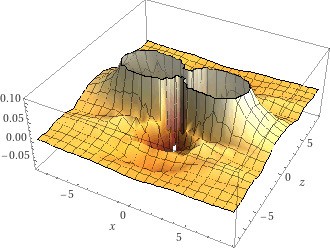}
         \caption{(a)}
         \label{plasmons1}
     \end{subfigure}

     \begin{subfigure}[b]{0.35\textwidth}
         \includegraphics[width=\textwidth]{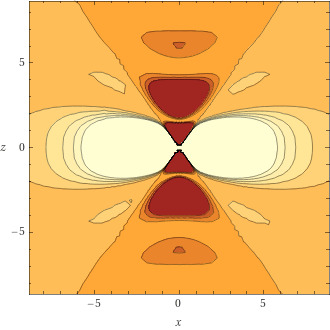}
         \caption{(b)}
         \label{plasmons2}
     \end{subfigure}
     \captionof{figure}{(a) 2D plot of the EM intensity distribution (in a.u.) below and above the device surface, $\omega/\nu_{em}=3.1$; (b) Contour plot of the EM intensity distribution at $y=0$.}
\end{figure}

\renewcommand{\thefigure}{4}
\begin{figure}
     \begin{subfigure}[b]{0.35\textwidth}
         \includegraphics[width=\textwidth]{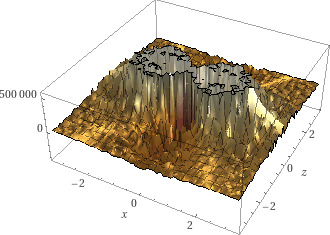}
         \caption{(a)}
         \label{plasmons3}
     \end{subfigure}

     \begin{subfigure}[b]{0.35\textwidth}
         \includegraphics[width=\textwidth]{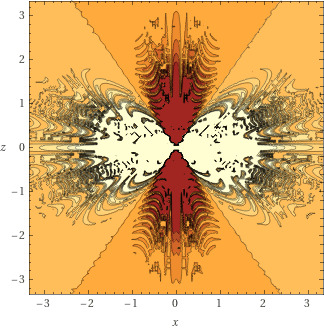}
         \caption{(b)}
         \label{plasmons4}
     \end{subfigure}
    \captionof{figure}{(a) 2D plot of the EM intensity distribution (in a.u.) below and above the device surface, $\omega/\nu_{em}=10^5$; (b) Contour plot of the EM intensity distribution at $y=0$.}
\end{figure}

In the natural world, a multifunctional, angle-dependent anti-reflection structure occurs on cicada wings, see Fig.~\ref{Fig2}. The insect's wings are composed of highly ordered, tiny vertical "nano-nipple" arrays, forming a biomorphic TiO$_2$ surface with small spaces between the ordered nano-antireflective structures allowing multiple reflective and scattering effects of the antireflective structures~\cite{Zada}. Another recent technique revealing the role of surface charges on cloaking and shielding was proposed using electroosmotic dipole flow that occurs around a localized surface charge domain under the application of an external electric field in a Hele-Shaw cell, revealing that the superposition of surface charge spots does produce complex flow patterns, without the application of physical walls~\cite{Paratore_2019}.

In stealth technology, we also may use a programmable plasmonic circuit was proposed in Ref.~\cite{Tu_2020} using a transparent patterned zinc oxide gate to provide full control of plasmons in graphene. These techniques may assist in better design stealth technology~\cite{Kim_2017} and Fig.~\ref{Fig3} exemplifies the concept. As wide spectrum surveillance (wide bandwidth capability) is needed because the frequency of the enemy radar is not known beforehand, the parameter $\nu_{em}$ may be adjusted to lock with the enemy frequency radar $\omega$~\cite{polymers}.

\subsection{Conclusion}

The above scaling and methodology, based on the hydrodynamic construct of the electromagnetic field, introduces a new tool to assess electromagnetic turbulence or practical means to control optical phenomena such as cloaking and shielding and further reveals the diffraction nature of EM turbulence. The diffraction pattern is dependent not only on the form of the encountered barriers but, additionally, is essentially linked to the surface density of surface plasmons vortices and electrons being on the obstacle. 

\section*{Competing Financial Interests}
This research received no specific grant from any funding agency in the public, commercial, or not-for-profit sectors.

\bibliographystyle{unsrtnat}


\end{document}